\documentclass[11pt]{article}
\begin{document}

\title{ 
{\bf Selected topics in integrable models}} 
\author{Ashok Das \\
\\
Department of Physics and Astronomy, \\
University of Rochester,\\
Rochester, New York, 14627-0171}
\date{}
\maketitle

\begin{abstract}

In these talks, I discuss a few selected topics in integrable models that
are of interest from various points of view. Some open questions are
also described. 
\end{abstract}

\vfill\eject
\section{Introduction:}

The subject of integrable models now encompasses a very large area of
research involving many seemingly different topics. It is not at all
possible to give a detailed exposition of the subject in just a few
lectures. Therefore, when the organizers of the Andr\'{e} Swieca
summer school asked me to choose a few topics on which to speak at the
school, I agreed with a lot of trepidation. In order to make the
lectures self-complete, I, of course, had to start with some
basics. The subsequent topics that I talked about, naturally, represent
a personal choice according to my interests. There are many other interesting
areas that are being pursued by many active groups, but I simply could
not have done justice to all, in the limited time available. Similarly,
the literature on the subject is vast and it would have been totally
impossible for me to even pretend to have a complete list of
references. Consequently, I have only chosen a handful of references,
for my talks, that I have absolutely used in the preparation of my
lectures. I apologize to all whose works I have not been able to mention
in my talks or references that I have not been able to list.

\section{Historical development:}

Let me begin with some introduction to the historical development of
the subject. Let us begin with the simple, working  definition of an
integrable model as a physical system that is described by nonlinear partial
differential equations, which can be exactly solved. We will make the
definition more precise as we go along. There are quite a few systems
of equations of this kind that arise in various physical theories. For
example, in $0+1$ dimensions, the Toda lattice is described by the set
of equations
\begin{eqnarray}
\dot{Q}_{i} & = & P_{i},\qquad i = 1,2,\cdots , N\nonumber\\
\dot{P}_{1} & = & - e^{-(Q_{2}-Q_{1})}\nonumber\\
\dot{P}_{N} & = & e^{-(Q_{N}-Q_{N-1})}\nonumber\\
\dot{P}_{\alpha} & = & e^{-(Q_{\alpha} - Q_{\alpha -1})} -
e^{-(Q_{\alpha +1} - Q_{\alpha})},\qquad \alpha = 2,3,\cdots , N-1
\end{eqnarray}
and consists of a chain of $N$ particles on a one dimensional lattice
at the coordinates $Q_{i}$, with $P_{i}$ representing the conjugate
momenta. 

In $1+1$ dimensions, similarly, there is the celebrated KdV equation
described by
\begin{equation}
{\partial u(x,t)\over \partial t} = u {\partial u\over \partial x} +
{\partial^{3} u\over \partial x^{3}}
\end{equation}
where $u$, for example, may describe the height of a water wave from
the normal surface. The variables of the KdV equation can be scaled to
have arbitrary coefficients in front of all the terms. As a
consequence, 
another form in which the KdV equation is also known corresponds to
\begin{equation}
{\partial u\over \partial t} = 6 u {\partial u\over \partial x} +
{\partial^{3} u\over \partial x^{3}}
\end{equation}
In $1+1$
dimensions, there is also the non-linear Schr\"{o}dinger equation
described by
\begin{eqnarray}
{\partial\psi(x,t)\over \partial t} & = & -
{\partial^{2}\psi\over \partial x^{2}} - 2\kappa |\psi|^{2}
\psi\nonumber\\
{\partial\overline{\psi}(x,t)\over \partial t} & = &
{\partial^{2}\overline{\psi}\over \partial x^{2}} + 2\kappa |\psi|^{2}
\overline{\psi} 
\end{eqnarray}
where $\kappa$ is a constant measuring the strength of the nonlinear
interaction and can be both positive or negative corresponding to an
attractive or repulsive interaction. There are several other
integrable systems in $1+1$ dimensions, but these two are the ones
that have been widely studied. There are fewer integrable systems in
$2+1$ dimensions, which include the Kadomtsev-Petviashvelli (KP)
equations and the Davey-Stewartson (DS) equations.

One of the most important features that all integrable models have is
that they possess soliton solutions to the equations of
motion. Solitons are defined as localized, non-dispersive solutions
that maintain their shape even after being scattered. Historically, of
course, research in this area grew out of J. Scott Russel's
observation, in 1834, of a solitary wave travelling for miles
maintaining its shape. It was only in 1895 that Korteweg and de Vries
gave a mathematical description of such shallow water waves, which is
known as the KdV equation. Being nonlinear and difficult to solve,
these equations, however, did not generate a lot of interest. In 1965,
Kruskal and Zabusky undertook a \lq\lq computer'' experiment, namely,
they wanted to numerically study the evolution of the solutions of the
KdV equation. What they found was impressive, namely, when certain
solutions of the KdV equation were scattered off each other, they
maintained their shape even after going through the scattering
region. Kruskal coined the term \lq\lq solitons'' for such solutions
in 1969 and the presence of such solutions generated an enormous
interest in such systems from then on. 

\noindent {\bf Non-dispersive solutions:}

Most physical linear equations have dispersive solutions and the
presence of non-dispersive solutions in such systems is quite
interesting. To appreciate the origin of such solutions and to see
their relation to the nonlinear interactions of the theory, let us
analyze the KdV equation,
\begin{equation}
{\partial u\over \partial t} = u {\partial u\over \partial x} +
{\partial^{3} u\over \partial x^{3}}
\end{equation}
As we have noted, such a system describes shallow water waves, where
we can think of $u(x,t)$ as representing the height of the water wave
from the normal surface of water. The first term on the right hand
side represents the nonlinear term. Let us, for a moment, look at the
KdV equation without the nonlinear term, namely,
\begin{equation}
{\partial u\over \partial t} = {\partial^{3} u\over \partial x^{3}}
\end{equation}
It is easy to write down the dispersion relation following from this
equation, 
\begin{equation}
E(k) = k^{3}
\end{equation}
This immediately tells us that the phase and the group velocities,
associated with a wave packet, in this case, are different, namely,
\begin{equation}
v_{\rm phase} = {E(k)\over k} = k^{2},\qquad v_{\rm group} = {d
E(k)\over dk} = 3k^{2}
\end{equation}
Thus, we see that if the KdV equation contained only the linear term
on the right hand side, solutions will disperse.

On the other hand, let us next assume that the KdV equation does not
contain  the linear term on the right hand side, namely,
\begin{equation}
{\partial u\over \partial t} = u {\partial u\over \partial x}
\end{equation}
This is also known as the Riemann equation. This can be solved by the
method of characteristics and the solution has the general form
\begin{equation}
u(x,t) = f(x + u t)
\end{equation}
which is quite interesting, for it says that the velocity of
propagation is directly proportional to the height of the
wave. Namely, the higher points of the wave will travel faster than
those at a lower height. This is what leads to the breaking of waves
etc. However, from our point of view, we see that this has a
localizing effect,  opposite of  what the linear
term leads to. The linear and the nonlinear terms, on the right hand
side of the KdV equation, therefore, have opposing behavior and if
they can balance each other exactly, then, we can have solutions that
will travel without any dispersion. The presence of nonlinear
interactions, therefore, is quite crucial to the existence of
non-dispersive solutions.

In the KdV equation, this indeed happens and we have non-dispersive
solutions. For example, let us consider
\begin{equation}
u(x,t) = 3 v {\rm sech}^{2} {\sqrt{v}\over 2} (x + vt)
\end{equation}
This gives
\begin{eqnarray*}
{\partial u\over \partial t} & = & -3 v^{5\over 2} {\rm sech}^{2}
{\sqrt{v}\over 2} (x + vt) \tanh {\sqrt{v}\over 2} (x + vt)\\
u {\partial u\over \partial x} & = & - 9 v^{5\over 2} {\rm sech}^{4}
{\sqrt{v}\over 2} (x + vt) \tanh {\sqrt{v}\over 2} (x + vt)\\
{\partial^{3} u\over \partial x^{3}} & = & - 3 v^{5\over 2} {\rm sech}^{2}
{\sqrt{v}\over 2} (x + vt) \tanh^{3} {\sqrt{v}\over 2} (x + vt) + 6
v^{5\over 2} {\rm sech}^{4} {\sqrt{v}\over 2} (x + vt) \tanh {\sqrt{v}\over
2} (x + vt)
\end{eqnarray*}
With trigonometric identities, it is easy to check now that
$$
{\partial u\over \partial t} = u {\partial u\over \partial x} +
{\partial^{3} u\over \partial x^{3}}
$$
holds so that this represents a solution of the KdV equation. From the
form of the solution, it is clear that it maintains its shape as it
travels (non-dispersive) and such a solution is known as a one soliton
solution. One can also construct multi-soliton solutions through what
is known as Backl\"{u}nd transformations, which I will not go into.

\noindent {\bf Conserved charges:}

Given the KdV equation, one can immediately construct three conserved
charges, namely, it is easy to check that
\begin{eqnarray}
H_{1} & = & \int dx\, u\nonumber\\
H_{2} & = & {1\over 2} \int dx\, u^{2}\nonumber\\
H_{3} & = & \int dx\,\left({1\over 3!} u^{3} - {1\over 2}
\left({\partial u\over \partial x}\right)^{2}\right)
\end{eqnarray}
are conserved under the evolution of the KdV equation. Several people
had also constructed up to 13 conserved charges for the system when
Kruskal conjectured that the KdV system has an infinite number of
functionally independent conserved charges. This was subsequently
proved and the conserved charges constructed through the Miura
transformation (as well as through the method of inverse
scattering). However, in retrospect, the presence of an infinite
number of conserved charges associated with a system possessing
soliton solutions is intuitively quite clear. As we have noted,
solitons scatter through each other maintaining their shape. This
implies that there must be conservation laws which prevents the
solution from deformations. Since the soliton is an extended solution,
there must, therefore, be an infinity of such conservation laws for
the solution to maintain its shape through collisions. The important
thing is that such a system has an infinite number of conserved
charges, which also means that the system is integrable.

\noindent{\bf Bi-Hamiltonian structure:}

Integrable systems are Hamiltonian systems. For example, in the case
of the KdV equation, we note that if we define
\begin{equation}
\{u(x) , u(y) \}_{1} = \partial \delta (x-y) = {\partial \over \partial x}
\delta (x-y)
\end{equation}
then, the KdV equation can be written in the Hamiltonian form as
\begin{equation}
{\partial u\over \partial t} = \{ u (x) , H_{3}\}_{1} = {\partial\over
\partial x} \left({1\over 2} u^{2} + {\partial^{2} u\over \partial
x^{2}}\right) = u {\partial u\over \partial x} + {\partial^{3} u\over
\partial x^{3}}
\end{equation} 
However, what is even more interesting is the fact that the same set
of equations can also be written in the Hamiltonian form if we define
\begin{equation}
\{ u(x) , u(y) \}_{2} = \left({\partial^{3}\over \partial x^{3}} +
{1\over 3} \left({\partial\over \partial x} u + u {\partial\over
\partial x}\right)\right) \delta (x-y)
\end{equation}
so that
\begin{equation}
{\partial u\over \partial t} = \{ u(x) , H_{2}\}_{2} =
\left({\partial^{3}\over \partial x^{3}} + {1\over
3}\left({\partial\over \partial x} u + u {\partial\over \partial
x}\right)\right) u = u {\partial u\over \partial x} + {\partial^{3}
u\over \partial x^{3}}
\end{equation}
Namely, the KdV equation is Hamiltonian with respect to at least two
distinct Hamiltonian structures.

Representing the two Hamiltonian structures as (operators acting on
delta function)
\begin{eqnarray}
{\cal D}_{1} & = & \partial\nonumber\\
{\cal D}_{2} & = & \partial^{3} + {1\over 3} (\partial u + u \partial)
\end{eqnarray}
we note that the first structure is what is normally called the
Abelian current algebra, while the second structure is known as the
Virasoro algebra. As a result, we do not have to worry about these
structures satisfying the Jacobi identity, which they do. This is
another general feature of integrable models, namely, the Hamiltonian
structures of integrable models are generally associated with
symmetry algebras. In fact, some of the nonlinear algebras, such as
the $W$ algebras, were studied from the point of view of integrable
models. 

Looking at the structure of the two Hamiltonian structures of the KdV
equation, it is clear that not only does this system have two
distinct Hamiltonian structures, but that that
\begin{equation}
{\cal D} = {\cal D}_{2} + \alpha {\cal D}_{1}
\end{equation}
also represents a genuine Hamiltonian structure (not necessarily of
the system). This is a nontrivial statement, considering that a
structure must satisfy the Jacobi identity - a nontrivial relation -
in order to qualify as a Hamiltonian structure. In this case, it
follows from the fact that ${\cal D}_{2}$ is a genuine Hamiltonian
structure for any variable $u$ and that
\begin{equation}
{\cal D}_{2} (u + {3\over 2} \alpha) = {\cal D}_{2} (u) + \alpha {\cal
D}_{1}
\end{equation}
When two Hamiltonian structures have such a relation (namely, if two
structures are Hamiltonian, then, a linear combination of the two is
also), they are said to be compatible. When a Hamiltonian system can
be described by two distinct Hamiltonian structures that are
compatible, the system is said to be a bi-Hamiltonian system. 

The existence of two Hamiltonian descriptions for the same equation,
of course, implies that
\begin{equation}
{\partial u\over \partial t} = {\cal D}_{2} {\delta H_{2}\over \delta
u} = {\cal D}_{1} {\delta H_{3}\over \delta u}
\end{equation}
This is a prototype of the recursion relation that exists between
conserved charges in such systems. One can define a recursion operator
\begin{equation}
{\cal R} = {\cal D}_{1}^{-1} {\cal D}_{2}
\end{equation}
which will relate the successive conserved charges as
\begin{equation}
{\delta H_{n+1}\over \delta u} = {\cal R} {\delta H_{n}\over \delta u}
\end{equation}
Furthermore, if the two Hamiltonian structures are compatible, one can
further show that these conserved charges are also in involution with
respect to either of the Hamiltonian structures, thereby proving that
the system is integrable.

The phase space geometry of integrable systems is quite
interesting. These are, of course, symplectic manifolds, but because
there are at least two distinct Hamiltonian structures (symplectic
structures), these are very special symplectic manifolds. Let us call
the two symplectic structures as $\omega_{1}$ and $\omega_{2}$. Then,
on this manifold, one can naturally define a nontrivial $(1,1)$ tensor
as
\begin{equation}
S = \omega_{1}^{-1} \omega_{2}
\end{equation}
The evolution of this equation can be thought of as the Lax equation
and, therefore, this gives a geometrical meaning to the Lax
equation. Furthermore, one can show that if the Nijenhuis torsion
tensor, associated with this $(1,1)$ tensor, vanishes, then, the
conserved charges will be in involution. Consequently, the vanishing
of the Nijenhuis torsion tensor can be thought of as a sufficient
condition for integrability in this geometrical description.  

Let me also note here that since the KdV system has an infinite number
of conserved quantities $H_{n}, n=1,2,\cdots $, each of this can be
thought of as a Hamiltonian and will lead to a flow as
\begin{equation}
{\partial u\over \partial t_{k}} = {\cal D}_{1} {\delta H_{k+1}\over
\delta u} = {\cal D}_{2} {\delta H_{k}\over \delta u}
\end{equation}
Thus, with every integrable system is a hierarchy of flows and these
represent the higher order flows of the system. The entire hierarchy
of flows shares the same infinite set of conserved quantities and are
integrable. 

\noindent{\bf Initial value problem:}

An interesting question, in connection with these nonlinear integrable
systems, is how can one solve the initial value problem. Namely, given
the initial values of the dynamical variables, in such systems, how
does one determine their values at any later time. In linear systems,
we are familiar with techniques such as the Fourier transformation or
the Laplace transformation, which help by transforming differential
equations into algebraic ones. However, these methods are not very
useful in dealing with nonlinear equations. The method that is useful
(and, therefore, which can be thought of as the analog of the Fourier
transformation in the case of nonlinear equations) is the method of
inverse scattering. Let me explain this in some detail.

Let us consider the linear Schr\"{o}dinger equation
\begin{equation}
\left(\partial^{2} + {1\over 6} u (x,t)\right) \psi = \lambda \psi
\end{equation}
where $\partial$ stands for ${\partial\over \partial x}$ and $u(x,t)$
is the dynamical variable of the KdV equation. Here $t$ is just a
parameter that the potential in the Schr\"{o}dinger equation depends
on and not the evolution parameter of the Schr\"{o}dinger
equation. Since the potential $u(x,t)$ depends on the parameter $t$,
it follows that both $\psi$ and $\lambda$ will depend on $t$ as
well. However, what Gardner, Greene, Kruskal and Miura observed was
that, if $u(x,t)$ satisfied the KdV equation, then, the eigenvalues,
$\lambda$, were independent of $t$, namely, the evolution is
isospectral in such a case, or,
\begin{equation}
\lambda_{t} = 0
\end{equation}
Furthermore, in such a case, the dependence of $\psi$ on $t$ is very
simple, namely,
\begin{equation}
\psi_{t} = - \left({1\over 6} u_{x} + \alpha\right) \psi +
\left(4\lambda + {1\over 3} u\right) \psi_{x}
\end{equation}

In such case, the evolution of the scattering data, such as the
reflection coefficient, the transmission coefficient etc, with $t$ is
easy to determine and, in fact, take a simple form. Thus, the strategy
for solving the initial value problem can be taken as follows. Let us
choose the linear Schr\"{o}dinger equation with the potential $u
(x,0)$ and determine the scattering data. Next determine the
scattering data at an arbitrary value of $t$ from the simple evolution
of the scattering data. Once we have the scattering data for an
arbitrary $t$, we can ask what is the potential, $u (x,t)$, which would give
rise to those scattering data. This is essentially the method of
inverse scattering. The reconstruction of the potential from the
scattering data is done through the Gel'fand-Levitan-Marchenko
equation,
\begin{equation}
K(x,y) + B(x,y) + \int_{x}^{\infty} dz\,K(x,z) B(y+z) = 0,\qquad y\geq
x
\end{equation}
where
\begin{equation}
B(x) = {1\over 2\pi} \int_{-\infty}^{\infty} dk\,R(k) e^{ikx} +
\sum_{n=1}^{N} c_{n} e^{-\kappa_{n} x}
\end{equation}
Here, $R$ is the coefficient of reflection, $\kappa_{n}$'s represent the
eigenvalues for the bound states and $c_{n}$'s correspond to the
normalization constants for the bound state wave functions. Once the
solution of the Gel'fand-Levitan-Marchenko equation is known, the
potential is determined from
\begin{equation}
{1\over 6} u(x) = 2 {\partial K(x,x)\over \partial x}
\end{equation}

Let us see explicitly how the method works, in an example. However,
let me also note that, while one can, in principle, find a solution to
the GLM equation, in practice it may be difficult unless the starting
potential were very special. One such class of potentials are
solitonic potentials which are known to be reflectionless. In such a
case,
\begin{equation}
R = 0
\end{equation}
and this makes calculations much simpler.

Let us, therefore, consider
\begin{equation}
u (x,0) = 12 {\rm sech}^{2} x
\end{equation}
We recognize, from our earlier discussion, that this is the one
soliton solution of the KdV equation. Such a potential leads to no
reflection. It supports only one bound state, for which
\begin{equation}
\kappa = 1,\qquad \psi (x,0) = {1\over 2} {\rm sech} x
\end{equation}
Therefore, we obtain
\begin{equation}
c(0) = \left(\int_{-\infty}^{\infty} dx\,\psi^{2} (x, 0)\right)^{-1} =
2
\end{equation}
From the equation for the \lq\lq time'' evolution of the wave
function, it is easy to determine that
\begin{equation}
c(t) = c(0) e^{-8t} = 2 e^{-8t}
\end{equation}
so that we have
\begin{equation}
B(x,t) = c(t) e^{-x} = 2 e^{-8t - x}
\end{equation}
In this case, the GLM equation becomes,
\begin{equation}
K(x,y,t) + 2 e^{-8t - x - y} + 2 \int_{x}^{\infty} dz\,K(x,z,t) e^{-8t
-y -z} = 0
\end{equation}
This determines
\begin{eqnarray}
K (x,y,t) & = & - {2e^{-8t - x - y}\over 1 + e^{-8t - 2x}}\nonumber\\
{\rm or,}\quad u (x,t) & = & 12 {\partial K(x,x)\over \partial x} = 12
{\rm sech}^{2} (x + 4t)
\end{eqnarray}
This is, of course, the solution that we had determined earlier
(corresponding to a specific choice of $v$) and this explains how the
method of inverse scattering works. (It is worth noting here that the
inverse scattering method was also independently used by Faddeev and
Zakharov to solve the KdV equation.)

\noindent{\bf The Lax equation:}

In some sense, the Lax equation is a formal generalization of the
ideas of Gardener, Greene, Kruskal and Miura. Let us consider a linear
operator, $L(t)$ that depends on a parameter $t$ through the
potential. Let us assume the eigenvalue equation
\begin{equation}
L (t) \psi = \lambda \psi
\end{equation}
along with the evolution of the wave function
\begin{equation}
{\partial \psi\over \partial t} = B \psi
\end{equation}
where $B$ represents an anti-symmetric operator. It follows now that
\begin{eqnarray}
{\partial L(t)\over \partial t} \psi + L(t) {\partial \psi\over
\partial t} & = & \lambda_{t} \psi + \lambda {\partial \psi\over
\partial t}\nonumber\\
{\rm or,}\quad {\partial L\over \partial t} \psi + L B \psi & = &
\lambda_{t} \psi + \lambda B \psi = \lambda_{t} \psi + B L
\psi\nonumber\\
{\rm or,}\quad {\partial L\over \partial t}\,\psi & = & \left[B ,
L\right]\,\psi + \lambda_{t} \psi
\end{eqnarray}
It follows, therefore, that
\begin{equation}
\lambda_{t} = 0
\end{equation}
provided
\begin{equation}
{\partial L\over \partial t} = \left[ B , L \right]
\end{equation}

This is known as the Lax equation and $L, B$ are called the Lax
pair. What this equation says is that the evolution of the linear
equation with respect to the parameter $t$ will be isospectral,
provided the Lax equation is satisfied ($\lambda$ is commonly referred
to as the spectral parameter.). Furthermore, if a Lax pair is
found such that the Lax equation yields a given nonlinear equation,
then, this says that one can associate a linear Schr\"{o}dinger
equation with it and the method of inverse scattering can be carried
out for this system leading to the integrability of the nonlinear
system. This is the power of the Lax equation and, as we have
mentioned earlier, the geometrical meaning of the Lax equation is that
it represents the evolution of a special $(1,1)$ tensor in the
phase space (symplectic manifold) of the system.

As an example, let us analyze the KdV equation in some detail. Let us
note that if we choose,
\begin{eqnarray}
L (t) & = & \partial^{2} + {1\over 6} u (x,t)\nonumber\\
B (t) & = & 4 \partial^{3} + {1\over 2} \left(\partial u + u
\partial\right)
\end{eqnarray}
then, with some straight forward computation, we can determine that
\begin{equation}
\left[ B , L \right] = {1\over 6} \left(u {\partial u\over \partial x}
+ {\partial^{3} u\over \partial x^{3}}\right)
\end{equation}
so that the Lax equation
$$
{\partial L\over \partial t} = \left[ B , L \right]
$$
leads to
\begin{eqnarray}
{1\over 6} {\partial u\over \partial t} & = & {1\over 6} \left( u
{\partial u\over \partial x} + {\partial^{3} u\over \partial
x^{3}}\right)\nonumber\\
{\rm or,}\quad {\partial u\over \partial t} & = & u {\partial u\over
\partial x} + {\partial^{3} u\over \partial x^{3}}
\end{eqnarray}
We recognize this to be the KdV equation and, having a Lax
representation for the equation, then, immediately determines the
linear Schr\"{o}dinger equation associated with it, which we have
described earlier, in connection with the method of inverse
scattering. This is, however, a general procedure that applies to any
integrable model and that is why the Lax equation plays an important
role in the study of integrable systems.
\vfill\eject

\noindent{\bf Zero curvature formalism:}

There is an alternate method of representing integrable systems, which
brings out some other properties associated with the system quite
nicely. Let us continue with the example of the KdV equation and
consider the following vector potentials.
\begin{eqnarray}
\bar{A}_{0} & = & \left(\begin{array}{cc}
{1\over 6} C_{x} + {\sqrt{\lambda}\over 3} C & - {1\over 6} C_{xx} -
{1\over 18} u C - {\sqrt{\lambda}\over 3} C_{x}\\
{1\over 3} C & - {1\over 6} C_{x} - {\sqrt{\lambda}\over 3} C
\end{array}\right)\nonumber\\
\bar{A}_{1} & = & \left(\begin{array}{cc}
\sqrt{\lambda} & - {1\over 6} u\\
1 & - \sqrt{\lambda}
\end{array}\right)
\end{eqnarray}
There are several things to note here. First, $C = C[u,\lambda]$ and
that the vector potentials belong to the Lie algebra of $SL(2,{\bf
R})$, namely, $\bar{A}_{0},\bar{A}_{1}\in SL(2,{\bf R})$.

The curvature (field strength) associated with these potentials can be easily
calculated and one recognizes that the vanishing of the curvature
yields the equations associated with the KdV hierarchy. This is seen
as follows.
\begin{equation}
F_{01} = \partial_{0}\bar{A}_{1} - \partial \bar{A}_{0} - \left[
\bar{A}_{0} , \bar{A}_{1} \right] = 0
\end{equation}
gives
\begin{equation}
u_{t} = C_{xxx} + {1\over 3} \left(\partial u + u \partial\right) C -
4 \lambda C_{x}
\end{equation}
For $\lambda = 0$ and $C = u$, this coincides with the KdV
equation. In general, we can expand in a power series of the form
\begin{equation}
C = \sum_{n=0}^{N} (4\lambda)^{N-n} C_{n}[u]
\end{equation} 
Substituting this into the equation and matching the corresponding
powers of $(4\lambda)$, we obtain
\begin{eqnarray}
C_{0} & = & 1\nonumber\\
\left(\partial^{3} + {1\over 3} \left(\partial u + u
\partial\right)\right) C_{n} & = & \partial C_{n+1},\qquad n =
0,1,2,\cdots , N-1\nonumber\\
u_{t} & = & \left(\partial^{3} + {1\over 3} \left(\partial u + u
\partial\right)\right) C_{N}
\end{eqnarray}
We recognize these as giving the recursion relation between the
conserved charges (which we have discussed earlier) as well as the $N$
th equation of the hierarchy. This is known as the zero curvature
representation of the integrable system and brings out the recursion
relation between the conserved charges, the current algebra etc quite
nicely.

\noindent{\bf Drinfeld-Sokolov formalism:}

Thus, we see that an integrable model can be represented as a scalar
Lax equation as well as a matrix zero curvature condition. The
natural question that arises is whether there is any connection
between the two.

To analyze this question, let us note that the scalar Lax equation for
KdV is described by the Lax pair which leads to the linear equations
\begin{eqnarray}
L \psi & = & \left(\partial^{2} + {1\over 6} u \right) \psi = \lambda
\psi\nonumber\\
{\partial \psi\over \partial t} & = & B \psi = \left(4 \partial^{3} +
{1\over 2} \left(\partial u + u \partial \right)\right)\psi
\end{eqnarray}
The scalar Lax equation can be thought of as the compatibility
condition for these two equations when the spectral parameter is
independent of $t$. We note that, while the second equation may appear
to  be third  order 
in the derivatives, with the use of the Schr\"{o}dinger equation, the
higher order derivatives can, in fact, be reduced. As a result, this
pair of equations appears to be at the most quadratic in the
derivatives. 

We know that a second order equation can be written in terms of two
first order equations. Keeping this in mind, let us define
\begin{equation}
\psi_{1} = (\partial \psi)
\end{equation} 
as well as a two component column matrix wavefunction
\begin{equation}
\Psi = \left(\begin{array}{c}
\psi_{1}\\
\psi
\end{array}\right)
\end{equation}
It is clear now that the linear Schr\"{o}dinger equation
$$
L \psi = \left(\partial^{2} + {1\over 6} u\right) \psi = \lambda \psi
$$
can be written in the matrix form as
\begin{equation}
\partial_{x} \Psi = A_{1} \Psi = \left(\begin{array}{cc}
0 & \lambda - {1\over 6} u\\
1 & 0
\end{array}\right) \Psi
\end{equation}
Similarly, the time evolution equation (depending on $B$) can also be
written as a matrix equation of the form
\begin{equation}
\partial_{t} \Psi = A_{0} \Psi
\end{equation}
The compatibility of these two matrix linear equations leads to the
zero curvature condition
\begin{equation}
\partial_{t} A_{1} - \partial_{x} A_{0} - \left[ A_{0} , A_{1}\right]
= 0
\end{equation}
These potentials, however, do not resemble the potentials that we
studied earlier. However, it is easy to check that the two sets of
potentials are related by a global ($\lambda$ is a constant
independent of $t$) similarity transformation. For example, note that
\begin{equation}
\bar{A}_{1} = S^{-1} A_{1} S,\qquad S = \left(\begin{array}{cc}
1 & - \sqrt{\lambda}\\
0 & 1
\end{array}\right)
\end{equation}
This, therefore, establishes the connection between the two formalisms
and tells us how to go from one to the other or {\em vice versa}.

\section{Pseudo-differential operators:}

With the basics of the previous section, we are now ready to discuss
some of the integrable models in some detail. The first thing that we
note is that the Lax formalism and the Lax pair is quite crucial in
the study of integrable models. However, finding a Lax pair, for a
given integrable model, seems like a formidable task. This is where
the Gel'fand-Dikii formalism comes to rescue.

Let us consider a general operator of the form
\begin{equation}
P = \sum_{i} a_{i} \partial^{i}
\end{equation}
where
\begin{equation}
\partial = {\partial\over \partial x},\qquad a_{i} = a_{i} (x)
\end{equation}
If $i\geq 0$, namely, if the operator $P$ only contains non-negative
powers of $\partial$, then, it is a differential operator ($i=0$ term
is a multiplicative operator) On the other hand, if $i$ takes also
negative values, then, the operator $P$ is known as a
pseudo-differential operator (Formally, $\partial^{-1}$ is defined
from $\partial \partial^{-1} = 1 = \partial^{-1}\partial$). There are
some standard nomenclature in using pseudo-differential
operators. Thus, for example,
\begin{equation}
P_{+} = \left(\sum_{i} a_{i} \partial^{i}\right)_{i \geq 0} = \cdots +
a_{1} \partial + a_{0}
\end{equation}
and, correspondingly,
\begin{equation}
P_{-} = \left(\sum_{i} a_{i} \partial^{i}\right)_{i < 0} = a_{1}
\partial^{-1} + a_{2} \partial^{-2} + \cdots
\end{equation}
By construction, therefore, we have
\begin{equation}
P = P_{+} + P_{-}
\end{equation}
We can also define, in a corresponding manner,
\begin{equation}
\left(P\right)_{\geq k} = \left(\sum_{i} a_{i}
\partial^{i}\right)_{i\geq k}
\end{equation}

Let us now note the standard properties of the derivative operator,
namely,
\begin{eqnarray}
\partial^{i} \partial^{j} & = & \partial^{i+j}\nonumber\\
\partial^{i} f & = & \sum_{k=0}^{\infty} \left(\begin{array}{c}
i\\
k
\end{array}\right) f^{(k)} \partial^{i-k}
\end{eqnarray}
which holds true for any $i,j$, positive or negative, where
\begin{equation}
\left(\begin{array}{c}
i\\
k
\end{array}\right) = {i (i-1) (i-2) \cdots (i-k+1)\over k!},\qquad
\left(\begin{array}{c}
i\\
0
\end{array}\right) = 1
\end{equation}
and $f^{(k)}$ denotes the $k$ th derivative of the function $f$. It is
worth noting here from the above formulae that positive powers of the
derivative operator acting to the right cannot give rise to negative
powers of the derivative and {\em vice versa}.

Using these properties of the derivative operators, we note that we
can define a multiplication of pseudo-differential operators. The
product of two pseudo-differential operators defines a
pseudo-differential operator and
 they define an algebra. We can define a residue of a
pseudo-differential operator to be the coefficient of $\partial^{-1}$,
in analogy with the standard residue, namely,
\begin{equation}
{\rm Residue}\quad P = {\rm Res}\quad P = a_{-1} (x)
\end{equation}
This allows us to define a concept called the trace of a
pseudo-differential operator as
\begin{equation}
{\rm Trace}\quad P = {\rm Tr}\quad P = \int dx\,{\rm Res}\quad P =
\int dx\, a_{-1}(x)
\end{equation}
Let us note that, given two pseudo-differential operators, $P,P'$,
\begin{equation}
{\rm Res}\quad \left[P , P'\right] = \left(\partial f(x)\right)
\end{equation}
In other words, the residue of the commutator of any two arbitrary
pseudo-differential operators is a total derivative. This, therefore,
immediately leads to the fact that
\begin{equation}
{\rm Tr}\quad PP' = {\rm Tr}\quad P'P
\end{equation}
since the \lq\lq trace'' of the commutator would vanish with the usual
assumptions on asymptotic fall off of variables. This shows that the
\lq\lq Trace'' defined earlier satisfies the usual cyclicity
properties and justifies the name.

Let us also note that, given a pseudo-differential operator
\begin{equation}
P = \sum_{i} a_{i} \partial^{i}
\end{equation}
we can define a dual operator as
\begin{equation}
Q = \sum_{i} \partial^{-i} q_{-i}
\end{equation}
where, the $q_{i}$'s are independent of the $a_{i}$'s. This allows us
to define a linear functional of the form
\begin{equation}
F_{Q} (P) = {\rm Tr}\quad PQ = \int dx\,\sum_{i} a_{i} q_{i-1}
\end{equation}

The Lax operators, as we have seen earlier in the case of the KdV
equation, have the form of differential operators. However, in
general, they can be pseudo-differential operators. Thus, there exist
two classes of Lax operators. Operators of the form
\begin{equation}
L_{n} = \partial^{n} + u_{1} \partial^{n-1} + u_{2} \partial^{n-2} +
\cdots + u_{n}
\end{equation}
are differential operators and lead to a description of integrable
models called the generalized KdV hierarchy. On the other hand, Lax
operators of the form
\begin{equation}
\Lambda_{n} = \partial^{n} + u_{1} \partial^{n-1} + \cdots + u_{n} +
u_{n+1} \partial^{-1} + \cdots
\end{equation}
correspond to pseudo-differential operators and lead to a description
of integrable models, commonly called the generalized KP hierarchy.

Let us consider the Lax operator for the generalized KdV hierarchy,
for the moment. Thus,
\begin{equation}
L_{n} = \partial^{n} + u_{1} \partial^{n-1} + \cdots + u_{n}
\end{equation}
We can now formally define the $n$ th root of this operator as
a general pseudo-differential operator of the form
\begin{equation}
\left(L_{n}\right)^{1\over n} = \partial + \sum_{i=0}^{\infty}
\alpha_{i}(x) \partial^{-i}
\end{equation}
such that
\begin{equation}
\left(L_{n}^{1\over n}\right)^{n} = L_{n}
\end{equation}
This allows us to determine all the coefficient functions,
$\alpha_{i}(x)$, iteratively and, therefore, the $n$ th root of the
Lax operator.

Let us next note that, since
\begin{equation}
\left[ L_{n}^{k\over n} , L_{n} \right] = 0
\end{equation}
for any $k$, it follows that
\begin{equation}
{\partial L_{n}\over \partial t_{k}} = \left[\left(L_{n}^{k\over
n}\right)_{+} , L_{n} \right] = - \left[ \left(L_{n}^{k\over
n}\right)_{-} , L_{n} \right], \qquad k \neq mn
\end{equation} defines a consistent Lax equation. This can be seen as
follows. First, if $k = mn$, then, 
$$
\left(L_{n}^{k\over n}\right)_{+} = \left(L_{n}^{m}\right)_{+} =
L_{n}^{m}
$$ and, therefore, the commutator will vanish and we will not have a
meaningful dynamical equation. For $k\neq mn$, we note, from the
structure of the first commutator, that it will, in general, involve
powers  of the derivative of the forms
$$
\partial^{n+k -1}, \partial^{n+k-2},\cdots , \partial^{0}
$$
On the other hand, the terms in the second commutator will, in
general, have  powers of the derivative of the forms
$$
\partial^{n-2}, \partial^{n-3}, \partial^{n-4},\cdots
$$
However, if the two expressions have to be equal, then, they can only
have nontrivial powers of the derivative of the forms
$$
\partial^{n-2}, \partial^{n-3}, \cdots , \partial^{0}
$$
This is precisely the structure of the Lax operator (except for the
term with $\partial^{n-1}$), which says that the above equation
represents a consistent Lax equation. This equation also will imply
that
$$
{\partial u_{1}\over \partial t} = 0
$$
which is why, often, this constant is set to zero (as is the case in,
say, the KdV equation). This result is very interesting, for once we
have a Lax operator, the other member of the pair can now be
identified with
\begin{equation}
B_{k} = \left(L_{n}^{k\over n}\right)_{+}
\end{equation}
up to a multiplicative constant. Such a Lax representation of a
dynamical system is known as the standard representation.

Furthermore, we note that the Lax equation also implies that
\begin{equation}
{\partial L_{n}^{1\over n}\over \partial t_{k}} = \left[
\left(L_{n}^{k\over n}\right)_{+} , L_{n}^{1\over n} \right]
\end{equation}
It is straight forward to show, using this, that
\begin{equation}
\partial_{t_{k}} \partial_{t_{m}} L_{n} = \partial_{t_{m}}
\partial_{t_{k}} L_{n}
\end{equation}
Namely, different flows commute. This is equivalent to saying that the
different Hamiltonians corresponding to the different flows are in
involution. Therefore, if we have the right number of conserved
charges, the system is integrable.

The construction of the conserved charges, therefore, is crucial in
this approach. However, we note from
\begin{equation}
{\partial L_{n}^{1\over n}\over \partial t_{k}} = \left[
\left(L_{n}^{k\over n}\right)_{+} , L_{n}^{1\over n} \right]
\end{equation}
that
\begin{equation}
{\partial\over \partial t_{k}} {\rm Tr}\quad L_{n}^{1\over n} = {\rm
Tr}\quad \left[
\left(L_{n}^{k\over n}\right)_{+} , L_{n}^{1\over n} \right] = 0
\end{equation}
which follows from the cyclicity of the \lq\lq trace''. Therefore, we
can identify the conserved quantities of the system (up to
multiplicative factors) with
\begin{equation}
H_{m} = {n\over m} {\rm Tr}\quad \left(L_{n}^{m\over n}\right),\qquad
m \neq ln
\end{equation}
This naturally gives the infinite number of conserved charges of the
system, which, as we have shown before, are in involution. Therefore,
in this description, integrability is more or less automatic.
\vfill\eject

\noindent{\bf Hamiltonian structures:} 

In the Lax formalism, we can also determine the Hamiltonian structures
of the system in a natural manner. These are known in the subject as
the Gel'fand-Dikii brackets and they are determined from the
observation that the Lax equation looks very much like Hamilton's
equation, with $\left(L_{n}^{k\over n}\right)_{+}$ playing the role of
the Hamiltonian and the commutator substituting for the Hamiltonian
structure. Analyzing this further, one ends up with two definitions of
Gel'fand-Dikii brackets, which give rise to the two Hamiltonian
structures of the system. With the notation of the linear functional
defined earlier, they can be written as
\begin{eqnarray}
\{F_{Q}(L_{n}) , F_{V}(L_{n})\}_{1} & = & {\rm Tr}\quad \left(L_{n}
\left[V , Q \right]\right)\nonumber\\
\{F_{Q}(L_{n}) , F_{V}(L_{n}) \}_{2} & = & {\rm Tr}\quad
\left(L_{n}Q\left(L_{n}V\right)_{+} - QL_{n}
\left(VL_{n}\right)_{+}\right)
\end{eqnarray}
The second bracket is particularly tricky if the Lax operator has a
constrained structure and the modifications, in such a case, are well
known and I will not get into that.

It is worth noting that these brackets are, by definition,
anti-symmetric as a Hamiltonian structure should be. While the first
bracket is manifestly anti-symmetric, the second is not. However, it
is easy to see that the second is also anti-symmetric in the following
way.
\begin{eqnarray}
\{F_{Q}(L_{n}) , F_{V}(L_{n})\} & = & {\rm Tr}\quad
\left(L_{n}Q\left(L_{n}V\right)_{+} -
QL_{n}\left(VL_{n}\right)_{+}\right)\nonumber\\
 & = & {\rm Tr}\quad \left(L_{n}QL_{n}V -
L_{n}Q\left(L_{n}V\right)_{-} - QL_{n}VL_{n} +
QL_{n}\left(VL_{n}\right)_{-}\right)\nonumber\\
 & = & {\rm Tr}\quad
\left(-\left(L_{n}Q\right)_{+}\left(- L_{n}V\right)_{-} +
\left(QL_{n}\right)_{+}\left(VL_{n}\right)_{-}\right)\nonumber\\
 & = & {\rm Tr}\quad \left(-\left(L_{n}Q\right)_{+}L_{n}V +
\left(QL_{n}\right)_{+}VL_{n}\right)\nonumber\\
 & = & - {\rm Tr}\quad \left(L_{n}V\left(L_{n}Q\right)_{+} -
VL_{n}\left(QL_{n}\right)_{+}\right)\nonumber\\
 & = & - \{F_{V}(L_{n}) , F_{Q}(L_{n})\}
\end{eqnarray}
Thus, the two brackets indeed satisfy the necessary ant-symmetry
property of Hamiltonian structures. Furthermore, it can also be shown
(I will not go into the details) that these brackets satisfy Jacobi
identity as well and, therefore, constitute two Hamiltonian structures
of the system.

Without going into details, I would like to make some general remarks
about the Lax operators of the KP type. Let us consider a Lax operator
of the type
\begin{equation}
\Lambda_{n} = \partial^{n} + u_{1} \partial^{n-1} + \cdots + u_{n} +
u_{n+1} \partial^{-1} + \cdots
\end{equation}
This is a pseudo-differential operator, unlike the earlier case, and,
as we have already remarked, such Lax operators describe generalized
KP hierarchies. In this case, it can be shown that a Lax equation of the form
\begin{equation}
{\partial \Lambda_{n}\over \partial t_{k}} =
\left[\left(\Lambda_{n}^{k\over n}\right)_{\geq m} ,
\Lambda_{n}\right]
\end{equation}
is consistent, only for $m=0,1,2$. This is, therefore, different from
the generalized KdV hierarchy that we have already studied. For $m
=0$, the Lax equation is called, as before, a standard representation,
while for $m=1,2$, it is known as a non-standard representation. All
the ideas that we had developed for the standard representation go
through for the non-standard representation as well and we will return
to such an example later.

\noindent{\bf Example:}

As an application of these ideas, let us analyze some of the
integrable models from this point of view. First, let us consider the
KdV hierarchy. In this case, we have already seen that
\begin{equation}
L = L_{2} = \partial^{2} + {1\over 6} u
\end{equation}
As we had noted earlier, we note that the coefficient of the linear
power of $\partial$ has been set to zero (which is consistent with the
Lax equation). In this case, we can determine the square root of the
Lax operator, following the method described earlier, and it has the
form,
\begin{equation}
L^{1\over 2} = \partial + {1\over 12} u \partial^{-1} - {1\over 24}
u_{x} \partial^{-2} + {1\over 48} \left(u_{xx} - {1\over 6}
u^{2}\right) \partial^{-3} + \cdots
\end{equation}
It is easy to check that the square of this operator leads to $L$ up
to the particular order of terms.

In this case, we have
\begin{equation}
\left(L^{1\over 2}\right)_{+} = \partial
\end{equation}
which gives
\begin{eqnarray}
{\partial L\over \partial t_{1}} & = & \left[\left(L^{1\over
2}\right)_{+} , L\right] = \left[ \partial , L\right]\nonumber\\
{\rm or,}\quad {\partial u\over \partial t_{1}} & = & {\partial u\over
\partial x}
\end{eqnarray}
This is the chiral boson equation and is known to be the lowest order
equation of the KdV hierarchy. Let us also note that
\begin{equation}
\left(L^{3\over 2}\right)_{+} = \left(LL^{1\over 2}\right)_{+} =
\partial^{3} + {1\over 4} u\partial + {1\over 8} u_{x} = \partial^{3}
+ {1\over 8} (\partial u + u \partial) = {1\over 4} B
\end{equation}
where $B$ is the second member of the Lax pair for the KdV equation
that we had talked about earlier. It is clear, therefore, that
\begin{equation}
{\partial L\over \partial t} = 4 \left[\left(L^{3\over 2}\right)_{+} ,
L\right]
\end{equation}
will lead to the KdV equation. Similarly, one can derive the higher
order equations of the KdV hierarchy from the higher fractional powers
of the Lax operator.

We note from the structure of the square root of $L$ that
\begin{equation}
{\rm Tr}\quad L^{1\over 2} = {1\over 12} \int dx\, u(x)
\end{equation}
Similarly,
\begin{eqnarray}
{\rm Tr}\quad L^{3\over 2} & = & \int dx\,\left[{1\over
48}\left(u_{xx}-{1\over 6} u^{2}\right) + {1\over 72}
u^{2}\right]\nonumber\\
 & = & {1\over 96} \int dx\, u^{2}(x)
\end{eqnarray}
Up to multiplicative constants, these are the first two conserved
quantities of the KdV hierarchy and the higher order ones can be
obtained similarly from the \lq\lq trace'' of higher fractional powers
of the Lax operator.

From the form of the Lax operator, in this case,
$$
L = \partial^{2} + {1\over 6} u
$$
we note that we can define the dual operators
\begin{equation}
Q = \partial^{-2}q_{2} + \partial^{-1}q_{1},\qquad V =
\partial^{-2}v_{2} + \partial^{-1}v_{1}
\end{equation}
Here $q_{i}$ and $v_{i}$ are supposed to be independent of the
dynamical variable $u$, so that the linear functionals take the forms
\begin{equation}
F_{Q}(L) = {\rm Tr}\quad LQ = {1\over 6}\int dx\,uq_{1},\qquad
F_{V}(L) = {1\over 6} \int dx\,uv_{1}
\end{equation}
In this case, we can work out
\begin{equation}
\{F_{Q}(L) , F_{V}(L)\}_{1} = {1\over 36} \int dx\
dy\,q_{1}(x)v_{1}(y) \{u(x) , u(y)\}_{1}
\end{equation}
On the other hand,
\begin{equation}
{\rm Tr}\quad L \left[V , Q\right] = \int dx\, \left(q_{1}v_{1,x} -
q_{1,x}v_{1}\right) = - 2\int dx\, q_{1,x} v_{1}
\end{equation}
Thus, comparing the two expressions, we obtain
\begin{equation}
\{u(x) , u(y)\}_{1} = 72 {\partial\over \partial x}\delta (x-y)
\end{equation}
We recognize this to be the correct first Hamiltonian structure for
the KdV equation (except for a multiplicative factor). The derivation
of the  second Hamiltonian structure
is slightly more involved since the structure of the KdV Lax operator
has a constrained structure (the linear power of $\partial$ is
missing). However, the construction through the Gel'fand-Dikii
brackets, keeping this in mind, can be carried through and gives the
correct second Hamiltonian structure for the theory.  

Let me note in closing this section that the generalization of the
method of  inverse scattering as well as the generalization of the Lax
formalism (or the Gel'fand-Dikii formalism) to higher dimensions is
not as well understood and remain open questions.

\section{Two boson hierarchy:}

In this section, I will describe another integrable system in $1+1$
dimensions, which is very interesting. The study of this system is of
fundamental importance, since this system can reduce to many others
under appropriate limit/reduction. It is described in terms of two
dynamical variables and has the form
\begin{eqnarray}
{\partial u\over \partial t} & = & \left(2h + u^{2} - \alpha
u_{x}\right)_{x}\nonumber\\
{\partial h\over \partial t} & = & \left(2uh + \alpha h_{x}\right)_{x}
\end{eqnarray}
Here $\alpha$ is an arbitrary constant parameter and we can think of
$h$ as describing the height of a water wave from the surface, while
$u$ describes the horizontal velocity of the wave. This equation,
therefore, describes general shallow water waves.

This system of equations is integrable and, as we have mentioned,
reduces to many other integrable systems under appropriate
limit/reduction. To name a few, let us note that, when the parameter
$\alpha = 0$, this system reduces to Benney's equations, which also
represents the standard, dispersionless long water wave equation. For
$\alpha = -1$ and $h = 0$, this gives us the Burger's equation. When
$\alpha =1$ and we identify
$$
u = - {q_{x}\over q},\qquad h = \bar{q}q
$$
we obtain the non-linear Schr\"{o}dinger equation from the two boson
equations. Similarly, both the KdV and the mKdV equations are
contained in this system as higher order flows. Thus, the study of this
system is interesting because once we understand this
system,,properties of all these systems are also known.

Conventionally, the two boson equation is represented with the
identifications
\begin{equation}
\alpha = 1,\qquad u = J_{0},\qquad h = J_{1}
\end{equation}
which is the notation we will follow in the subsequent discussions. In
these notations, therefore, the two boson equations take the form
\begin{eqnarray}
{\partial J_{0}\over \partial t} & = & \left(2J_{1} + J_{0}^{2} -
J_{0,x}\right)_{x}\nonumber\\
{\partial J_{1}\over \partial t} & = & \left(2J_{0}J_{1} +
J_{1,x}\right)_{x}
\end{eqnarray}

Being integrable, this system of equations can be described by a Lax
equation. Let us consider the Lax operator
\begin{equation}
L = \partial - J_{0} + \partial^{-1} J_{1}
\end{equation}
so that it is a pseudo-differential operator. Let us note that, for
this operator,
\begin{eqnarray}
\left(L^{2}\right)_{\geq 1} & = & \partial^{2} - 2 J_{0}
\partial\nonumber\\
\left(L^{3}\right)_{\geq 1} & = & \partial^{3} - 3J_{0} \partial^{2} +
3 (J_{1} + J_{0}^{2} - J_{0,x}) \partial
\end{eqnarray}
and so on. It is now straight forward to compute and show that
\begin{equation}
\left[ L , \left(L^{2}\right)_{\geq 1} \right] =  - \left(2J_{1} +
J_{0}^{2} - J_{0,x}\right)_{x} + \partial^{-1} \left(2J_{0}J_{1} +
J_{1,x}\right)
\end{equation}
It is, therefore, clear that the non-standard Lax equation
\begin{equation}
{\partial L\over \partial t} = \left[ L , \left(L^{2}\right)_{\geq
1}\right]
\end{equation}
gives as consistent equations the two boson equations.

The higher order flows of the two boson (TB) hierarchy can be obtained
from
\begin{equation}
{\partial L\over \partial t_{k}} = \left[ L , \left(L^{k}\right)_{\geq
1}\right]
\end{equation}
Of particular interest to us is the next higher order equation coming
from
\begin{equation}
{\partial L\over \partial t_{3}} = \left[ L , \left(L^{3}\right)_{\geq
1}\right]
\end{equation}
These have the forms
\begin{eqnarray}
{\partial J_{0}\over \partial t} & = & - J_{0,xxx} -
3\left(J_{0}J_{0,x}\right)_{x} - 6 \left(J_{0}J_{1}\right)_{x} -
\left(J_{0}^{3}\right)_{x}\nonumber\\
{\partial J_{1}\over \partial t} & = & - \left(J_{1,xx} + 3
\left(J_{0}J_{1}\right)_{x} + 3 \left(J_{1}(J_{1} + J_{0}^{2} -
J_{0,x})\right)\right)_{x}
\end{eqnarray}
This equation is interesting, for we note that if we set $J_{0}=0$ and
identify $J_{1}={1\over 6} u$, the equations reduce to the KdV
equation. That is, as we had mentioned earlier, the KdV equation is
contained in the higher flows of the TB hierarchy. Furthermore, it is
also interesting to note that this provides a nonstandard
representation of the KdV equation, unlike the earlier example where
it was described by a standard Lax equation.

Given the Lax representation, the conserved charges are easily
constructed by the standard procedure from
\begin{equation}
H_{n} = {\rm Tr}\quad L^{n}
\end{equation}
so that we have
\begin{eqnarray}
H_{1} & = & {\rm Tr}\quad L = \int dx\,J_{1}\nonumber\\
H_{2} & = & {\rm Tr}\quad L^{2} = \int dx\, J_{0}J_{1}\nonumber\\
H_{3} & = & {\rm Tr}\quad L^{3} = \int dx\, \left(J_{1}^{2} -
J_{0,x}J_{1} + J_{1} J_{0}^{2}\right)
\end{eqnarray}
and so on. It is clear that if we set $J_{0}=0$ and identify $J_{1} =
{1\over 6} u$ all the even conserved charges vanish and the odd ones
coincide with the conserved charges of the KdV equation.

\noindent{\bf Hamiltonian structures:}

Let us denote the generic Hamiltonian structure associated with this
system as 
\begin{equation}
\left(\begin{array}{cc}
\{J_{0},J_{0}\} & \{J_{0},J_{1}\}\\
\{J_{1},J_{0}\} & \{J_{1},J_{1}\}
\end{array}\right) = {\cal D} \delta (x-y)
\end{equation}
Then, it can be shown that the TB equation has three Hamiltonian
structures. But, let me only point out the first two here.
\begin{eqnarray}
{\cal D}_{1} & = & \left(\begin{array}{cc}
0 & \partial\\
\partial & 0
\end{array}\right)\nonumber\\
{\cal D}_{2} & = & \left(\begin{array}{cc}
2\partial & \partial J_{0} - \partial^{2}\\
J_{0}\partial + \partial^{2} & \partial J_{1} + J_{1} \partial
\end{array}\right)
\end{eqnarray}
so that we can write the two boson equations as 
\begin{equation}
\partial_{t} \left(\begin{array}{c}
J_{0}\\
J_{1}
\end{array}\right) = {\cal D}_{1} \left(\begin{array}{c}
{\delta H_{3}\over \delta J_{0}}\\
{\delta H_{3}\over \delta J_{1}}
\end{array}\right) = {\cal D}_{2} \left(\begin{array}{c}
{\delta H_{2}\over \delta J_{0}}\\
{\delta H_{2}\over \delta J_{1}}
\end{array}\right)
\end{equation}
It is now easily checked that, under $J_{0}\rightarrow J_{0} +
\alpha$, where $\alpha$ is an arbitrary constant, 
\begin{equation}
{\cal D}_{2} \rightarrow {\cal D}_{2} + \alpha {\cal D}_{1}
\end{equation}
which proves that these two Hamiltonian structures are compatible and
that the system is integrable (which we already know from the Lax
description of the system).

Normally, the second Hamiltonian structure of an integrable system is
related to some symmetry algebra. To see this connection, in this
case, let us redefine (this is also known as changing the basis)
\begin{equation}
J(x) = J_{0}(x),\qquad T(x) = J_{1} - {1\over 2} J_{0,x}(x)
\end{equation}
In terms of these new variables, the second Hamiltonian structure takes the
form
\begin{eqnarray}
\{J(x),J(y)\}_{2} & = & 2 \partial_{x} \delta (x-y)\nonumber\\
\{T(x),J(y)\}_{2} & = & J(x) \partial_{x} \delta (x-y)\nonumber\\
\{T(x),T(y)\}_{2} & = & \left(T(x) + T(y)\right)\partial_{x} \delta
(x-y) + {1\over 2} \partial_{x}^{3} \delta (x-y)
\end{eqnarray}
We recognize this as the Virasoro-Kac-Moody algebra, which is the
bosonic limit of the twisted $N=2$ superconformal algebra.
\vfill\eject

\noindent{\bf Non-linear Schr\"{o}dinger equation:}

Since the TB system reduces to the non-linear Schr\"{o}dinger
equation, we can also find a Lax description for that system from the
present one. We note that with the identification
\begin{equation}
J_{0} - {q_{x}|over q},\qquad J_{1} = \bar{q} q
\end{equation}
the Lax operator for the TB system becomes,
\begin{eqnarray}
L & = & \partial - J_{0} + \partial^{-1} J_{1}\nonumber\\
 & = & \partial + {q_{x}\over q} + \partial^{-1}\bar{q} q\nonumber\\
 & = & q^{-1} \left(\partial + q \partial^{-1} \bar{q}\right) q\nonumber\\
 & = & G \tilde{L} G^{-1}
\end{eqnarray}
where we have defined
\begin{equation}
G = q^{-1},\qquad \tilde{L} = \partial + q \partial^{-1} \bar{q}
\end{equation}
Namely, the two Lax operators $L$ and $\tilde{L}$ are related by a
gauge transformation. The adjoint of this transformed Lax operator is
determined to be
\begin{equation}
{\cal L} = \tilde{L}^{\dagger} = - \left(\partial + \bar{q}
\partial^{-1} q\right)
\end{equation}
It is straightforward to check that both the standard Lax equations
\begin{equation}
{\partial \tilde{L}\over \partial t} =
\left[\tilde{L},\left(\tilde{L}^{2}\right)_{+}\right]
\end{equation}
and 
\begin{equation}
{\partial {\cal L}\over \partial t} = \left[\left({\cal
L}^{2}\right)_{+},{\cal L}\right]
\end{equation}
give rise to the non-linear Schr\"{o}dinger equation. (However,
supersymmetry seems to prefer the second representation.)

Furthermore, if we identify
\begin{equation}
\bar{q} = q = u
\end{equation}
then, the standard Lax equations
\begin{equation}
{\partial \tilde{L}\over \partial t} =
\left[\tilde{L},\left(\tilde{L}^{3}\right)_{+}\right]
\end{equation}
and
\begin{equation}
{\partial {\cal L}\over \partial t} = \left[\left({\cal
L}^{3}\right)_{+},{\cal L}\right]
\end{equation}
give the mKdV equation.

Thus, we see that the TB system is indeed a rich theory to study.

\section{Supersymmetric equations:}

Given a supersymmetric integrable system, we can ask if there also
exist supersymmetric integrable systems corresponding to it. It turns
out to be a difficult question in the sense that the
supersymmetrization turns out not to be unique and we do not yet fully
understand how to classify all possible supersymmetrizations of such
systems. Let me explain this with an example.

Let us consider the KdV equation
$$
{\partial u\over \partial t} = 6 uu_{x} + u_{xxx}
$$
Then, a supersymmetric generalization of this system that is also
integrable is given by
\begin{eqnarray}
{\partial u\over \partial t} & = & 6 uu_{x} + u_{xxx} - 3 \psi
\psi_{xx}\nonumber\\
{\partial \psi\over \partial t} & = & 3 \left(u\psi\right)_{x} +
\psi_{xxx}
\end{eqnarray}
Here $\psi$ represents the fermionic superpartner of the bosonic
dynamical variable $u$ of the KdV equation. It is easy to check that
these equations remain invariant under the supersymmetry
transformations,
\begin{equation}
\delta \psi = \epsilon u,\qquad \delta u = \epsilon \psi_{x}
\end{equation}
where $\epsilon$ is a constant Grassmann parameter (fermionic
parameter) of the transformation, satisfying $\epsilon^{2} = 0$.

We can, of course, determine the supersymmetry charge (generator of
supersymmetry) associated with this system and it can be shown that
the supersymmetry algebra satisfied by this charge is
\begin{equation}
\left[Q , Q\right]_{+} = 2 P
\end{equation}
This can also be checked by taking two successive supersymmetry
transformations in opposite order and adding them. There are several
things to note from this system. First, the second Hamiltonian
structure associated with this system, ${\cal D}_{2}$, is the
superconformal algebra, which is the supersymmetrization of the
Virasoro algebra. Second, just as this represents a$N=1$
supersymmetric extension of the KdV equation, there also exists a
second $N=1$ supersymmetric extension,
\begin{eqnarray}
{\partial u\over \partial t} & = & 6 u u_{x} + u_{xxx}\nonumber\\
{\partial \psi\over \partial t} & = & 6 u \psi_{x} + \psi_{xxx}
\end{eqnarray}
which is integrable. This second supersymmetric extension was
originally discarded as being a \lq\lq trivial'' supersymmetrization,
since the bosonic equations do not change in the presence of
fermions. However, it generated a lot of interest after it was
realized that it is this equation that arises in a study of
superstring theory from the point of view of matrix models. Such a
supersymmetric extension now has the name -B
supersymmetrization. Thus, we see that even at the level of $N=1$
supersymmetrization, there is no unique extension of the integrable
model. The problem becomes more and more severe as we go to higher
supersymmetrizations. For $N=2$, it is known that there are at least
four distinct \lq\lq nontrivial'' supersymmetrizations of the KdV
equation that are integrable. Understanding how many distinct
supersymmetrizations are possible for a given integrable equation,
therefore, remains an open question. In addition, there are also
fermionic extensions of a given integrable model (not necessarily
supersymmetric) that are also integrable and there does not exist any
unified description of them yet.

\noindent{\bf Lax description:}

Since the supersymmetric KdV equation (super KdV) is an integrable
system, let us determine a Lax description for it. The simplest
way to look for a Lax description is to work on a superspace, which is
the natural manifold to study supersymmetric systems.

Let us consider a simple superspace parameterized by $(x,\theta)$, a
single bosonic coordinate $x$ and a single real fermionic coordinate
$\theta$, satisfying $\theta^{2} =0$. Supersymmetry, then, can be
shown to correspond to a translation, in this space, of the fermionic
coordinate  $\theta$. On this space, one can define a covariant
derivative
\begin{equation}
D = {\partial\over \partial \theta} + \theta {\partial\over \partial
x}
\end{equation}
which transforms covariantly under a supersymmetry transformation and
can be seen to satisfy
\begin{equation}
D^{2} = \partial
\end{equation}
On the superspace, a function is called a superfield and, since the
fermionic coordinate is nilpotent, has a simple representation of the
form (in the present case)
\begin{equation}
\Phi (x,\theta) = \psi (x) + \theta u(x)
\end{equation}
The Grassmann parity of the components is completely determined by the
parity of the superfield. For our discussion of the super KdV system,
let us choose the superfield $\Phi$ to be fermionic so that we can
think of $u$ as the bosonic dynamical variable of the KdV equation and
$\psi$ as its fermionic superpartner.

In terms of this superfield, the super KdV equations can be combined
into one single equation of the form
\begin{equation}
{\partial \Phi\over \partial t} = \left(D^{6}\Phi\right) + 3
\left(D^{2}\left(\Phi (D\Phi)\right)\right)
\end{equation}
It is now easy to check that if we choose, as Lax operator on this
superspace,
\begin{equation}
L = D^{4} + D \Phi
\end{equation}
then, the Lax equation
\begin{equation}
{\partial L\over \partial t} = \left[ \left(L^{3\over 2}\right)_{+} ,
L\right]
\end{equation}
gives the super KdV equation. The structure of this Lax operator and
the Lax equation is, of course, such that they reduce to the KdV
equation in the bosonic limit, which is nice.

It is worth making a few remarks about operators on the
superspace. First, a pseudo-differential operator on this space is
defined with powers of $D$. Correspondingly, various decompositions
are done with respect to powers of $D$. Thus, we define
\begin{eqnarray}
{\rm super\ Residue}\quad P & = & {\rm sRes}\quad P = {\rm
coefficient\ of}\,D^{-1}\nonumber\\
{\rm super\ Trace}\quad P & = & {\rm sTr}\quad P = \int dx\,d\theta\,
{\rm sRes}\quad P
\end{eqnarray}
The conserved quantities, for the super KdV system, for example, are
obtained as
\begin{equation}
H_{n} = {\rm sTr}\quad L^{2n+1\over 2}
\end{equation}
These are all bosonic conserved quantities and there is an infinite
number of them. They all reduce to the infinite set of 
conserved charges of the bosonic integrable model in the bosonic
limit. The Gel'fand-Dikii brackets can be generalized to this space as
well and lead to the correct Hamiltonian structures of the theory. An
interesting feature of the Lax description on a superspace is that the
same integrable system can be described in terms of a Lax operator
that is either bosonic or fermionic. Furthermore, in the
supersymmetric models we can define non-local conserved charges from
the Lax operator by taking, say, for example in the super KdV case,
powers of quartic roots of the Lax operator. Finally, let me also
point out that a supersymmetric -B system has fermionic Hamiltonians
(conserved charges) with corresponding odd (fermionic) Hamiltonian
structures.

\noindent{\bf Super TB hierarchy:}

As we have seen, the TB hierarchy consists of two dynamical variables,
$J_{0},J_{1}$. Therefore, the supersymmetrization of this system will
involve two fermionic partners, say $\psi_{0},\psi_{1}$. From our
experience with the super KdV system, let us combine them into two
fermionic superfields of the forms
\begin{equation}
\Phi_{0} = \psi_{0} + \theta J_{0},\qquad \Phi_{1} = \psi_{1} + \theta
J_{1}
\end{equation}
With a little bit of algebra, we can check that if we choose the Lax
operator
\begin{equation}
L = D^{2} - (D\Phi_{0}) + D^{-1} \Phi_{1}
\end{equation}
then, we obtain
\begin{equation}
\left(L^{2}\right)_{\geq 1} = D^{4} - 2 (D\Phi_{0}) D^{2} - 2 \Phi_{1}
D
\end{equation}
which gives
\begin{equation}
\left[L , \left(L^{2}\right)_{\geq 1}\right] =
\left(D\left((D^{4}\Phi_{0}) - D(D\Phi_{0})^{2} - 2
(D^{2}\Phi_{1})\right)\right) + D^{-1} \left((D^{4}\Phi_{1}) + 2
D^{2}(\Phi_{1}(D\Phi_{0}))\right)
\end{equation}
It is clear, therefore, that the Lax equation
\begin{equation}
{\partial L\over \partial t} = \left[L , \left(L^{2}\right)_{\geq
1}\right]
\end{equation}
leads to consistent equations and gives
\begin{eqnarray}
{\partial \Phi_{0}\over \partial t} & = & - (D^{4}\Phi_{0}) + 2
(D\Phi_{0})(D^{2}\Phi_{0}) + 2 (D^{2}\Phi_{1})\nonumber\\
{\partial \Phi_{1}\over \partial t} & = & (D^{4}\Phi_{1}) + 2
\left(D^{2}(\Phi_{1}(D\Phi_{0}))\right)
\end{eqnarray}
The Lax operator as well as the Lax equation (and the equations
following from it are easily seen to reduce to the TB equations in the
bosonic limit.  These, therefore, represent a supersymmetric extension
of the TB hierarchy that is integrable. The higher order flows of the
hierarchy are obtained from
\begin{equation}
{\partial L\over \partial t_{k}} = \left[ L , \left(L^{k}\right)_{\geq
1}\right]
\end{equation}

Once we have the Lax description of the system, we can immediately
construct the conserved charges from
\begin{equation}
H_{n} = {\rm sTr}\quad L^{n} = \int dx\,d\theta\, {\rm sRes}\quad
L^{n},\qquad n = 1,2,3,\cdots
\end{equation}
Explicitly, we can construct the lower order conserved charges as
\begin{eqnarray}
H_{1} & = & - \int dx\,d\theta\, \Phi_{1}\nonumber\\
H_{2} & = & 2 \int dx\,d\theta\, (D\Phi_{0})\Phi_{1}\nonumber\\
H_{3} & = & 3 \int dx\,d\theta\,\left((D^{3}\Phi_{0}) - (D\Phi_{1}) -
(D\Phi_{0})^{2}\right)\Phi_{1} 
\end{eqnarray}
and so on. Thus, we see that the system has an infinite number
conserved charges, which are in involution (follows from the Lax
description) and, therefore, is integrable.

\noindent{\bf Hamiltonian structures:}

Just like the bosonic system, the super TB equations also possess
three Hamiltonian structures, although they are not necessarily local
unlike the structures in the bosonic case. Let me only describe the
first two here.

Defining, as in the bosonic case, a generic Hamiltonian structure as,
\begin{equation}
\left(\begin{array}{cc}
\{\Phi_{0},\Phi_{0}\} & \{\Phi_{0},\Phi_{1}\}\\
\{\Phi_{1},\Phi_{0}\} & \{\Phi_{1},\Phi_{1}\}
\end{array}\right) = {\cal D} \delta (z - z') = {\cal D} \delta
(x-x')\delta (\theta - \theta')
\end{equation}
we note that
\begin{equation}
{\cal D}_{1} = \left(\begin{array}{cc}
0 & - D\\
-D & 0
\end{array}\right)
\end{equation}
as well as
\begin{equation}
{\cal D}_{2} = \left(\begin{array}{cc}
- 2D - 2D^{-1}\Phi_{1} D^{-1} + D^{-1} (D^{2}\Phi_{0}) D^{-1} & D^{3}
- D (D\Phi_{0}) + D^{-1} \Phi_{1} D\\
- D^{3} - (D\Phi_{0}) D - D \Phi_{1} D^{-1} & - D^{2} \Phi_{1} -
\Phi_{1} D^{2}
\end{array}\right)
\end{equation}
define the first two Hamiltonian structures of the super TB
hierarchy. These structures have the necessary symmetry property and
it can be checked using the method of prolongation that these
structures satisfy the Jacobi identity. They give rise to super TB
equation, say for example, through
\begin{equation}
\left(\begin{array}{c}
{\partial \Phi_{0}\over \partial t}\\
{\partial \Phi_{1}\over \partial t}
\end{array}\right) = {\cal D}_{1} \left(\begin{array}{c}
{\delta H_{3}\over \delta \Phi_{0}}\\
{\delta H_{3}\over \delta \Phi_{1}}
\end{array}\right) = {\cal D}_{2} \left(\begin{array}{c}
{\delta H_{2}\over \delta \Phi_{0}}\\
{\delta H_{2}\over \delta \Phi_{1}}
\end{array}\right)
\end{equation}

Although it is not obvious, the second Hamiltonian structure
corresponds to the twisted $N=2$ superconformal algebra, which can be
seen as follows. Let us look at the second Hamiltonian structure in
the component variables. Let us define
\begin{equation}
\xi = {1\over 2} \psi_{1},\qquad \bar{\xi} = - {1\over 2}
\left(\psi_{0,x} - \psi_{1}\right)
\end{equation}
In terms of these variables, the nontrivial elements of the second
Hamiltonian structure take the local form
\begin{eqnarray}
\{J_{0}(x),J_{0}(y)\}_{2} & = & 2 \partial_{x}\delta(x-y)\nonumber\\
\{J_{0}(x),J_{1}(y)\}_{2} & = &
\partial_{x}\left(J_{0}\delta(x-y)\right) -
\partial_{x}^{2}\delta(x-y)\nonumber\\
\{J_{1}(x),J_{1}(y)\}_{2} & = & (J_{1}(x)+J_{1}(y)) \partial_{x}\delta
(x-y)\nonumber\\
\{J_{0}(x),\xi (y)\}_{2} & = & \xi \delta (x-y)\nonumber\\
\{J_{0}(x),\bar{\xi} (y) \}_{2} & = & - \bar{\xi} \delta
(x-y)\nonumber\\
\{J_{1} (x), \xi (y)\}_{2} & = & (\xi (x) + \xi (y)) \partial_{x}
\delta (x-y)\nonumber\\
\{J_{1} (x), \bar{\xi} (y)\}_{2} & = & \bar{\xi}(x) \partial_{x}\delta
(x-y)\nonumber\\
\{\bar{\xi} (x), \xi (y)\}_{2} & = & -{1\over 4} J_{1} \delta (x-y) +
{1\over 4} \partial_{x}\left(J_{0}\delta (x-y)\right) - {1\over 4}
\partial_{x}^{2} \delta (x-y)
\end{eqnarray}
We recognize this to be the $N=2$ superconformal algebra.

\noindent{\bf ${\bf N=2}$ supersymmetry:}

Although the super TB system that we have constructed, naively appears
to have $N=1$ supersymmetry, in fact, it does possess a $N=2$
supersymmetry. This is already suggested by the fact that the second
Hamiltonian structure of this system corresponds to the $N=2$
superconformal algebra. Explicitly, this can be checked as follows.

Let us note that the super TB equations, in terms of the redefined
components, take the forms
\begin{eqnarray}
{\partial J_{0}\over \partial t} & = & - J_{0,xx} + 2 J_{0}J_{0,x} + 2
J_{1,x}\nonumber\\
{\partial J_{1}\over \partial t} & = & J_{1,xx} + 2
\left(J_{0}J_{1}\right)_{x} + 8
\left(\xi\bar{\xi}\right)_{x}\nonumber\\
{\partial \xi\over \partial t} & = & \xi_{xx} + 2 \left(\xi
J_{0}\right)_{x}\nonumber\\
{\partial \bar{\xi}\over \partial t} & = & \bar{\xi}_{xx} + 2
\left(\bar{\xi}J_{0}\right)_{x}
\end{eqnarray}
It is now straight forward to check that this system of equations is
invariant under the following two sets of supersymmetric transformations,
\begin{eqnarray}
\delta J_{0} & = & 2 \epsilon \xi\nonumber\\
\delta J_{1} & = & 2 \epsilon \xi_{x}\nonumber\\
\delta \xi & = & 0\nonumber\\
\delta \bar{\xi} & = & - {1\over 2} \epsilon \left(J_{0,x} -
J_{1}\right)
\end{eqnarray}
and
\begin{eqnarray}
\bar{\delta} J_{0} & = & 2 \bar{\epsilon} \bar{\xi}\nonumber\\
\bar{\delta} J_{1} & = & 0\nonumber\\
\bar{\delta} \xi & = & - {1\over 2} \bar{\epsilon} J_{1}\nonumber\\
\bar{\delta} \bar{\xi} & = & 0
\end{eqnarray}

\noindent{\bf Nonlocal charges:}

As we have already noted, in supersymmetric integrable systems, in
addition to the local bosonic conserved charges, we also have nonlocal
conserved charges. Let us study this a little within the context of
the super TB hierarchy.

Let us recall that the Lax operator for the system is given by
$$
L = D^{2} - (D\Phi_{0}) + D^{-1} \Phi_{1}
$$
and the infinite set of local, bosonic conserved charges are obtained
as
\begin{equation}
H_{n} = {\rm sTr}\quad L^{n},\qquad n=1,2,3, \cdots
\end{equation}
On the other hand, let us also note that in this system, we can also define
a second infinite set of conserved charges as 
\begin{equation}
Q_{2n-1\over 2} = {\rm sTr}\quad L^{2n-1\over 2},\qquad n = 1,2,3, \cdots
\end{equation}
There are several things to note from this definition. First of all,
unlike the earlier set, these conserved charges are fermionic in
nature. Second, they are nonlocal. Let me write down a few lower order
charges of this set.
\begin{eqnarray}
Q_{1\over 2} & = & - \int dx\,d\theta\, (D^{-1}\Phi_{1}) = - \int
dz\,(D^{-1}\Phi_{1}) \nonumber\\
Q_{3\over 2} & = & - \int dz\,\left[{3\over 2} (D^{-1}\Phi_{1})^{2} -
\Phi_{0}\Phi_{1} -
\left(D^{-1}((D\Phi_{0})\Phi_{1})\right)\right]\nonumber\\
Q_{5\over 2} & = & - \int dz\,\left[{1\over 6} (D^{-1}\Phi_{1})^{3} -
\left(5 (D^{-2}\Phi_{1})\Phi_{1} - 2\Phi_{0}\Phi_{1} - 3 (D\Phi_{1}) -
(D^{-1}\Phi_{1})^{2}\right) (D\Phi_{0})\right.\nonumber\\
 &  & \quad \left. + \left(D^{-1}\left((D\Phi_{1}) \Phi_{1} + \Phi_{1}
(D\Phi_{0})^{2} - (D\Phi_{1})(D^{2}\Phi_{0})\right)\right)\right]
\end{eqnarray}
and so on.

There are several things to be noted about these charges. First, as we
have already mentioned and as can be explicitly seen, these charges
are fermionic in nature and are conserved. Second, even though they
are defined on the superspace, they are not invariant under
supersymmetry (it is the nonlocality that is responsible for this
problem). This infinite set of charges satisfies an algebra which
appears to have the structure of a Yangian algebra, which arises in
the study of supersymmetric nonlinear sigma models, namely,
\begin{eqnarray}
\{Q_{1\over 2},Q_{1\over 2}\}_{1} & = & 0\nonumber\\
\{Q_{1\over 2},Q_{3\over 2}\}_{1} & = & H_{1}\nonumber\\
\{Q_{1\over 2},Q_{5\over 2}\}_{1} & = & H_{2}\nonumber\\
\{Q_{3\over 2},Q_{3\over 2}\}_{1} & = & 2 H_{2}\nonumber\\
\{Q_{3\over 2}, Q_{5\over 2}\}_{1} & = & {7\over 3} H_{3} + {7\over
24} H_{1}^{3}\nonumber\\
\{Q_{5\over 2},Q_{5\over 2}\}_{1} & = & 3 H_{4} - {5\over 8} H_{2}
H_{1}^{2}
\end{eqnarray}
and son on. The role of the fermionic nonlocal charges as well as the
meaning of the Yangian algebra, however, are not fully understood.

\section{Dispersionless integrable systems:}

Let us consider again the KdV equation as an example.
$$
{\partial u\over \partial t} = 6 u {\partial u\over \partial x} +
{\partial^{3} u\over \partial x^{3}}
$$
As we have noted earlier, it is the linear term on the right hand side
that is the source of dispersion for the solutions. Let us, therefore,
get rid of the dispersive term, in which case, the equation becomes
\begin{equation}
{\partial u\over \partial t} = 6 u {\partial u\over \partial x}
\end{equation}
This is known as the Riemann equation and we see that it corresponds
to the dispersionless limit of the KdV equation. Given a nonlinear,
bosonic equation, the systematic way in which the dispersionless limit
is obtained is by scaling
\begin{equation}
{\partial\over \partial t} \rightarrow \epsilon {\partial\over
\partial t},\qquad {\partial\over \partial x} \rightarrow \epsilon
{\partial \over \partial x}
\end{equation}
in the equation and then taking the limit $\epsilon\rightarrow
0$. This leads to the dispersionless limit of the original system of
equations. It is important to note here that the dynamical variable is
not scaled (although in supersymmetric systems, as we will see, it is
necessary to scale the fermionic variables to maintain supersymmetry).

Let us recall that the Lax description for the KdV equation is
obtained from the Lax operator
$$
L = \partial^{2} + u
$$
through the Lax equation
$$
{\partial L\over \partial t} = 4 \left[\left(L^{3\over 2}\right)_{+} ,
L\right]
$$
As we will now see, the Lax description for the dispersionless model
is obtained in a much simpler fashion. Let us consider a Lax function,
on the classical phase space, of the form
\begin{equation}
L (p) = p^{2} + u
\end{equation}
Here $p$ represents the classical momentum variable on the phase space
and, therefore, this Lax function only consists of commuting
quantities. However, we can think of this as consisting of a power
series in $p$ and formally calculate
\begin{equation}
\left(L^{3\over 2}\right)_{+} = p^{3} + {3\over 2} u p
\end{equation}
where the projections are defined with respect to the powers of
$p$. It is now simple to check, with the standard canonical Poisson
bracket relations,
$$
\{x,p\} = 1,\qquad \{x,x\} = 0 \{p,p\}
$$
that the Lax equation
\begin{equation}
{\partial L\over \partial t} = 4 \{L, \left(L^{3\over 2}\right)_{+}\}
\end{equation}
leads to the Riemann equation, which is the dispersionless limit of
the KdV equation.

Thus, we see that given a Lax description of an integrable model in
terms of Lax operators, the dispersionless limit is obtained from a
simpler Lax description on the classical phase space through classical
Poisson brackets. Let us also note that the conserved quantities of
the dispersionless model can be obtained from this Lax function as (in
this model)
\begin{equation}
H_{n} = {\rm Tr}\quad L^{2n+1\over 2} = \int dx, {\rm Res}\quad
L^{2n+1\over 2},\qquad n = 0,1,2,\cdots
\end{equation}
where \lq\lq Res'' is defined as the coefficient of the $p^{-1}$
term. All of our discussions in connection with pseudo-differential
operators carries through to this case where the Lax function has a
polynomial structure in the momentum variables. (When I gave these
talks, I had mentioned that the construction of the Hamiltonian
structure from the Gel'fand-Dikii was an open question. Since then,
this problem has been solved and we know now that these can be
constructed rather easily from a Moyal-Lax representation of
integrable models.)

Let me say here that dispersionless models encompasses a wide class of
systems such as hydrodynamic equations, polytropic gas dynamics,
Chaplygin gas, Born-Infeld equation, Monge-Amp\`{e}re equation,
elastic medium equation etc, some of which show up in the study of
string theory, membrane theory as well as in topological field
theories.

Let us study an example of such systems in some detail, namely, the
polytropic gas dynamics. These are described by a set of two equations
\begin{eqnarray}
u_{t} + u u_{x} + v^{\gamma -2} v_{x} & = & 0,\qquad \gamma \neq
0,1\nonumber\\
v_{t} + (u v)_{x} & = & 0
\end{eqnarray}
These equations are known to be Hamiltonian with
\begin{equation}
H = \int dx\,\left(-{1\over 2} u^{2}v - {v^{\gamma -1}\over \gamma
(\gamma -1)}\right)
\end{equation}
and
\begin{equation}
{\cal D} = \left(\begin{array}{cc}
0 & \partial\\
\partial & 0
\end{array}\right)
\end{equation}
so that we can write the polytropic gas equations as
\begin{equation}
\left(\begin{array}{c}
u_{t}\\
v_{t}
\end{array}\right) = {\cal D} \left(\begin{array}{c}
{\delta H\over \delta u}\\
{\delta H\over \delta v}
\end{array}\right)
\end{equation}
In fact, this system has three distinct Hamiltonian structures, but I
will not get into the details of this.

Let us next consider a Lax function, on the classical phase space, of
the form 
\begin{equation}
L = p^{\gamma -1} + u + {v^{\gamma -1}\over (\gamma -1)^{2}}
p^{-(\gamma -1)}
\end{equation}
Then, it is straight forward to check that the classical Lax equation
\begin{equation}
{\partial L\over \partial t} = {\gamma -1\over \gamma}
\{\left(L^{\gamma\over \gamma -1}\right)_{\geq 1}, L\}
\end{equation}
gives rise to the equations for the polytropic gas. The higher order
equations of the hierarchy are similarly obtained from
\begin{equation}
{\partial L\over \partial t_{n}} = c_{n} \{\left(L^{n+{1\over \gamma
-1}}\right)_{\geq 1}, L\}
\end{equation}
Thus, we see that the polytropic gas dynamics is obtained as a
nonstandard Lax description on the classical phase space. (At the time
of the school, this was the only Lax description for the system that
was known. Subsequently, a standard Lax description has been obtained,
which brings out some interesting connection between this system and
the Lucas polynomials.)

Once we have the Lax description, we can, of course, obtain the
conserved quantities from the \lq\lq Trace''. However, in this case,
unlike in the case of Lax operators, we observe an interesting
feature, namely, the residue can be obtained from expanding around $p
= 0$ or around $p = \infty$. Thus, there are two series of conserved
charges that we can construct for this system. Expanding around
$p=\infty$, we obtain
\begin{equation}
H_{n+1} = C_{n+1} {\rm Tr}\quad L^{n+1 - {1\over \gamma -1}},\qquad
n=0,1,2,\cdots
\end{equation} 
where $C_{n+1}$'s are normalization constants. Explicitly, the first
few of the charges have the forms
\begin{eqnarray}
H_{1} & = & \int dx\,u\nonumber\\
H_{2} & = & \int dx\, \left({1\over 2!} u^{2} + {v^{\gamma -1}\over
(\gamma -1)(\gamma -2)}\right)\nonumber\\
H_{3} & = & \int dx\,\left({1\over 3!} u^{3} + {uv^{\gamma -1}\over
(\gamma -1)(\gamma -2)}\right)
\end{eqnarray}
and so on. On the other hand, an expansion around $p=0$ leads to
\begin{equation}
\tilde{H}_{n} = \tilde{C}_{n} {\rm Tr}\quad L^{n + {1\over \gamma
-1}},\qquad n = 0,1,2,\cdots
\end{equation}
The first few charges of this set have the explicit forms,
\begin{eqnarray}
\tilde{H}_{0} & = & \int dx\, v\nonumber\\
\tilde{H}_{1} & = & \int dx\, uv\nonumber\\
\tilde{H}_{2} & = & \int dx\,\left({1\over 2!} u^{2}v +
{v^{\gamma}\over \gamma (\gamma -1)}\right)
\end{eqnarray}
and so on. The two sets of conserved quantities can, in fact, be
expressed in closed forms. Let us also note that if we define two
functions as
\begin{eqnarray}
\chi & = & \lambda^{-{1\over \gamma -1}}
\left\{\left[\left({u+\lambda\over 2}\right)^{2} - {v^{\gamma -1}\over
(\gamma -1)^{2}}\right]^{1\over 2} + {u+\lambda\over
2}\right\}^{1\over \gamma -1}\nonumber\\
 \tilde{\chi} & = & \lambda^{-{1\over \gamma -1}}
\left\{\left[\left({u+\lambda\over 2}\right)^{2} - {v^{\gamma -1}\over
(\gamma -1)^{2}}\right]^{1\over 2} - {u+\lambda\over
2}\right\}^{1\over \gamma -1}
\end{eqnarray}
where $\lambda$ is an arbitrary constant parameter, it can be shown
that these two functions generate the two sets of conserved quantities
as the coefficients of distinct powers of $\lambda$. Let me also note
that the second Hamiltonian structure for the polytropic gas has the
form
\begin{equation}
{\cal D}_{2} = \left(\begin{array}{cc}
\partial v^{\gamma -2} + v^{\gamma -2} \partial & \partial u + (\gamma
-2) u \partial\\
(\gamma -2) \partial u + u \partial & \partial v + v \partial
\end{array}\right)
\end{equation}
\vfill\eject

\noindent{\bf Dispersionless supersymmetric KdV:}

Let us recall that the super KdV equation can be described by the Lax
operator
$$
L = D^{4} + D \Phi
$$
and the Lax equation
$$
{\partial L\over \partial t} = 4 \left[\left(L^{3\over 2}\right)_{+} ,
L\right]
$$
Here,
$$
D = {\partial\over \partial \theta} + \theta {\partial\over \partial
x}
$$ represents the covariant derivatiove on the superspace. In trying
to obtain the dispersionless limit of this supersymmetric system, let
us recall what we have learnt from the dispersionless limit of a
bosonic model. We noted that the Lax operator goes over to the Lax
function with $\partial \rightarrow p$. However, our Lax operator, in
the supersymmetric case, is described in terms of super covariant
derivative $D$. Therefore, the natural question is what this object
goes over to in the dispersionless limit.

Let us note that the classical super phase space is parameterized by
$(x,\theta,p,p_{\theta})$ . From these we can define a variable
\begin{equation}
\Pi = - (p_{\theta} + \theta p)
\end{equation}
whose action on any phase space variable, through the Poisson brackets
is
\begin{equation}
\{\Pi , A\} = (DA),\qquad \{\Pi , \Pi\} = -2p
\end{equation}
Therefore, it would seem natural to let $D\rightarrow \Pi$ in the
dispaersionless limit. However, this leads to a serious problem. For
example, we know that $D^{2} = \partial\rightarrow p$ whereas $\Pi^{2}
= 0$ since it is a classical fermionic variable.

To analyze this problem a little more, let us recall that the
dispersionless limit is obtained by scaling
$$
\partial_{t}\rightarrow \epsilon \partial_{t},\qquad \partial
\rightarrow \epsilon \partial
$$
Therefore, since $D^{2} = \partial$, consistency would require that we
scale
\begin{equation}
D = {\partial\over \partial \theta} + \theta {\partial\over \partial
x}  \rightarrow \epsilon^{1\over 2} D
\end{equation}
This implies that the fermionic coordinate needs to be scaled as
\begin{equation}
\theta \rightarrow \epsilon^{-{1\over 2}} \theta
\end{equation}
On the other hand, let us recall that the basic fermionic superfield
of the theory is given as
\begin{equation}
\Phi (x,\theta) = \psi + \theta u
\end{equation}
Since the dynamical variable $u$ does not scale and $\theta$ scales,
it follows that supersymmetry can be maintained under such a scaling
only if
\begin{equation}
\psi \rightarrow \epsilon^{-{1\over 2}} \psi, \qquad \Phi\rightarrow
\epsilon^{-{1\over 2}} \Phi
\end{equation}
This shows that, unlike in the bsoonic theory, in a supersymmetric
theory, fermions scale in going to the dispersionless limit. With this
scaling, we can go to the super KdV equation and note that, in the
dispersionless limit, the equation becomes
\begin{equation}
\Phi_{t} = 3 D^{2} \left(\Phi (D\Phi)\right)
\end{equation}
which can be thought of as the super Riemann equation.

Obtaining a Lax function and, therefore, a Lax description of a
supersymmetric theory reamins an open question. However, through brute
force construction it is known that the Lax function
\begin{equation}
L = p^{2} + {1\over 2} (D\Phi) + {p^{-2}\over 16} \left((D\Phi)^{2} -
2\Phi \Phi_{x}\right) - {p^{-4}\over 32} \Phi (D\Phi) \Phi_{x}
\end{equation}
leads through the classical Lax equation (the projection is with
respect to powers of $p$)
\begin{equation}
{\partial L\over \partial t} = 4 \{L , \left(L^{3\over 2}\right)_{+}\}
\end{equation}
gives the dispersionless limit of the super KdV equation (super
Riemann equation). It is worth emphasizing here that, although the Lax
description for afew supersymmetric dispersionless models have been
constructed through brute force, a systematic understanding of them is
still lacking.

The conserved charges can be obtained from this Lax description
immediately. Thus,
\begin{equation}
H_{n} = C_{n} {\rm Tr}\quad L^{2n+1\over 2} = \int
dz\,\left(\Phi(D\Phi)^{n} - n\Phi (D\Phi)^{n-1} \Phi_{x}\right)
\end{equation}
where $n=0,1,2,\cdots$ and it is clear that these are bosonic
conserved charges. The supersymmetric dispersionless model has the
Hamiltonian structure
\begin{equation}
{\cal D} = -{1\over 2} \left(3 \Phi D^{2} + (D\Phi) D + 2
(D^{2}\Phi)\right)
\end{equation}
which can be recognized as the centerless superconformal algebra. With
this Hamiltonian structure, it is straight forward to check that the
conserved charges are in involution,
\begin{equation}
\{H_{n},H_{m}\} = \int dz\,{\delta H_{n}\over \delta \Phi} {\cal D}
{\delta H_{m}\over \delta \Phi} = 0
\end{equation}
which also follows from the Lax description of the system.

Let me note in closing that this supersymmetric model has two infinite
sets of nonlocal charges of the forms
\begin{eqnarray}
F_{n} & = & \int dz\, (D^{-1}\Phi)^{n}\nonumber\\
G_{n} & = & \int dz\,\Phi (D^{-1}\Phi)^{n}
\end{eqnarray}
where $n =1,2,\cdots$. (Note that $G_{0} = H_{0}$.) These charges have
been constructed by brute force, since it is not clear how to obtain
nonlocal quantities from a classical Lax function. This remains an
open question. Second, of the two sets of nonlocal charges, we see
that $F_{n}$ is fermionic while $G_{n}$ is bosonic. Furthermore, all
these conserved charges satisfy a very simple algebra,
\begin{eqnarray}
\{H_{n}, H_{m}\} & = & 0 = \{F_{n},H_{m}\} =
\{G_{n},H_{m}\}\nonumber\\
\{F_{n},G_{m}\} & = & 0 = \{G_{n},G_{m}\}\nonumber\\
\{F_{n},F_{m}\} & = & nm G_{n+m-2}
\end{eqnarray}

In connection with dispersionless supersymmetric integrable models,
several questions remain. For example, it is not clear how to
systematically construct the Lax description for them. It is not at
all clear how nonlocal charges can be obtained from a classical Lax
function. Neither is it clear what is the role played by these charges
within the context of integrability.

It is a pleasure to thank the organizers of the Swieca school for the
warm hospitality. This work was supoported in part by US DOE Grant
no. DE-FG-02-91ER40685.

\end{document}